\documentclass[11pt]{article}

\usepackage{graphicx}
\usepackage{color}
\usepackage{epsfig}

\newcommand{\tr}{^{\prime}}
\def \btheta{\mbox{\boldmath$\theta$}}
\def \bTheta{\mbox{\boldmath$\Theta$}}
\def \bzeta{\mbox{\boldmath$\zeta$}}
\def \boeta{\mbox{\boldmath$\eta$}}
\def\b#1{\mbox{\boldmath $#1$}}    
\def\bl#1{\mbox{\footnotesize \boldmath {$#1$}}} 
\def\cg#1{\mbox{${\cal #1}$}}

\renewcommand{\th}{\theta}

\newcommand{\Th}{\Theta}

\begin{document}
%

\title{A multidimensional latent class Rasch model for the assessment of the Health-related Quality of Life}
\author{Silvia Bacci\footnote{Dipartimento di Economia,
Finanza e Statistica, Universit\`a di Perugia, Via Pascoli 20, 06123
Perugia, Italy} \footnote{{\em email:}
silvia.bacci@stat.unipg.it}$\:$, Francesco
Bartolucci$^*$\footnote{{\em email:} bart@stat.unipg.it}} \maketitle



%

\emph{This work is a temporary and partial version. The final and integrated version has been accepted as contributed chapter in Christensen K.B., Kreiner S., Mesbah M. (eds.), Rasch related models and methods for health science, Wiley-ISTE, ISBN: 978-1-84821-222-0, expected for December 2012.}

\section{Introduction}
The World Health Organization $[$WHO 95$]$ defined the Quality of Life (QoL) as:
\begin{quote}
``\textit{the individuals' perceptions of their position in life in the
context of their culture and the value systems in which they live,
and in relation to their goals, expectations, standards and
concerns. It is a broad-ranging concept affected in a complex way
by the persons' physical health, psychological state, level of
independence, social relationships, personal beliefs, and their
relationship to the salient features of their environment.}''
\end{quote}
In a survey about quality of life, Fayers and Machin $[$FAY
00$]$ defined the Health related Quality of Life (HrQoL) as:
 \begin{quote}
``\textit{the way, which according to health of
a person influences his/her capacity
to lead on physical and social normal activities.}''
\end{quote}
The ``normality'' of an activity is a variable concept, 
where it
depends on the reference population. The use of indicators about
HrQoL in clinical and epidemiological contexts is various:
\textit{in primis} these indicators give additional information to
evaluate the effect of care therapies on a patient.
More and more, HrQoL is considered as a secondary
endpoint of clinical trials, because the main focus is on the
survival. However, in some contexts, such as that of pain therapies
for terminal cancer patients, it rises to a primary endpoint.

The main problem related to HrQoL concerns its measurement
because this characteristic is not directly observable.
Suitable measurement methods are then needed which
are typically based on qualitative information, coming from
\textit{ad hoc} questionnaires, to be translated
into quantitative information; see
$[$MES 10$]$ for a detailed illustration about statistical aspects
involved in the measurement of HrQoL. In such a way it is possible
to evaluate the patient's condition in relation to the general
condition of the population so as to provide
clinicians with information useful for the care and therapy
decisional process. In this context, the Rasch model $[$RAS 61$]$
represents an important tool to measure HrQoL.
However, some important aspects must be taken into
account.

First of all, HrQoL is a latent multidimensional concept, because
its proper evaluation requires consideration of several
dimensions (corresponding to different latent traits) that reflect
individual health conditions and how well patients are coping with
the stress due to illness. Often, two macro-dimensions can be
distinguished: a physical one and a psychological one; these are
further separable in sub-dimensions, such as bodily pain and physical
functioning in the former case, and mental health or vitality in the
latter one. Usually, the latent traits corresponding to the
different dimensions are highly correlated and are also correlated
with other latent traits, such as those corresponding to some
psychopathological disturbs, mainly anxiety and depression.
However, the classic Rasch model is based on the
assumption of unidimensionality. The easiest approach that is
adopted when this assumption is not realistic consists of estimating
separate Rasch models for subsets of items measuring different
latent traits, but this method does not allow us to measure the
correlation between these latent traits. A more suitable approach is based on in the multidimensional extension of Rasch models.

The second aspect is that, in many applications, it
is of interest to detect homogeneous classes of individuals who have
very similar latent characteristics. Detecting these classes of
individuals can be, not only more realistic, but also more
convenient for the decisional process because
individuals in the same class will receive the same clinical
treatment.

In order to analyse HrQoL data taking into account
the above aspects, in this chapter we proposed the use of a version
of the Rasch model which belongs to the class of multidimensional
Item Response Theory (IRT) models proposed by Bartolucci $[$BAR
07$]$. The model is characterized by two main features: (\textit{i})
more latent traits are simultaneously considered
(\textit{multidimensionality assumption}); (\textit{ii}) these
latent traits are represented by a random vector having a \textit{discrete}
distribution common to all subjects (\textit{discreteness
assumption}). Each support point of this distribution identifies a
different class of individuals. Obviously, these  are latent
classes, in the sense that we do not know to which class a given
individual belongs; moreover, we do not know how many latent classes
exist. The adopted model is then related to the latent class (LC)
model (see $[$LAZ 68$]$; $[$GOO 74$]$) and for this reason will be
referred to as the {\em multidimensional LC Rasch model}.

We note that the LC model originates as a method
to classify individuals on the basis of categorical responses, but,
more recently, the same discrete latent structure on which this model
is based has been exploited to account for the unobserved
heterogeneity between subjects into other models. Using this
structure can be considered as an alternative to the inclusion of
continuous random effects which avoids a parametric specification of
the distribution of these effects. A semi-parametric model then
results. In particular, this structure has been used by $[$LIN 91$]$
and $[$FOR 95$]$ to define a unidimensional LC Rasch model, which is
a particular case of the model that we here adopt. An alternative generalization of the Rasch model to the LC analysis is
represented by the mixture Rasch model of Rost $[$ROS 90$]$.
It may be seen as an extension of LC Rasch model, which allows for different
sets of item parameters for each latent class.

Another model which is strongly related to the
multidimensional LC Rasch model is the LC factor model proposed by
Magidson and Vermunt $[$MAG 01$]$. However, in the first approach
each item response is affected by only one of the latent traits, and
these latent traits may be correlated, whereas Magidson and Vermunt
$[$MAG 01$]$ assume that each item response may be simultaneously
affected by two or more latent traits, which are mutually
independent. We also have to mention alternative
specifications of a multidimensional Rasch model that have been
proposed; see, among others, the multidimensional marginally
sufficient Rasch model proposed by Hardouin and Mesbah $[$HAR
04$]$.

Aim of this chapter is also that of studying the correlation between
the latent dimensions of HrQoL in cancer patients and those behind
some psychopathological disturbs. This analysis
involves two tests of dimensionality. The first is a likelihood
ratio (LR) test which is based on the
multidimensional LC Rasch model 
that exploits the
discrete (or LC) marginal maximum likelihood (MML) approach. The
second test is based on the Martin-L\"{o}f test (ML) approach $[$MAR
70$]$ and exploits the conditional maximum likelihood (CML)
estimation method (see also $[$GLA 95b$]$). Alternative tests have
been proposed by several authors; for a review see Verhelst $[$VER
01$]$. In particular, we mention the approach of
Christensen et al. $[$CHR 02$]$, who propose a test similar to the
first test we here use, but it is based on the
assumption that the latent traits follow a multivariate
normal distribution.

The remainder of this chapter is structured as
follows. In Section 2 we describe the dataset used for the
illustrative application. In Section 3 we present the
multidimensional LC Rasch model of Bartolucci $[$BAR
07$]$, with special attention to model assumptions and estimation
from the practical point of view. In Section 4 we illustrate how
to estimate the correlation between latent traits on
the basis of the estimated parameters of this
model. In the same section we describe the two
tests of dimensionality based on the MML and CML approaches.
Finally, in Section 5 we present the results of the application to
the dataset described in Section 2.

\section{The dataset} \label{sec2}
In order to illustrate the approach presented in this chapter, we
analyze data which come from an Italian multi-centric clinical study.
These data concern $275$ oncological patients recruited from three different centres (Ancona, Perugia, and Messina).
Patients were asked to fill some questionnaires about different
latent characteristics; here, we consider HrQoL, anxiety, and
depression. In particular, HrQoL is assessed by the ``$36$-item
Short-Form Health Survey'' (SF-$36$) of Ware et al. $[$WAR 02$]$,
whereas anxiety and depression are assessed by the ``Hospital
Anxiety and Depression Scale'' (HADS) of Zigmond and Snaith $[$ZIG
83$]$. The response rate is equal to $74\%$ ($203$ patients out of
$275$). However, the sample of respondents is here assumed to be
representative of the entire sample, since there are
no significant differences between the distributions of age, gender, marital status,
education, and cancer diagnosis; see
Table~\ref{tab:descr} for a comparison between the population and
sample distributions of these variables.

\begin{table}[ht]\centering
\caption{\em Entire and respondent sample distributions of
age, gender, marital status, education, and cancer diagnosis (column
percentages).} \label{tab:descr} {\small
\begin{tabular}{lcc}
\hline\hline
    &   Entire sample  &    Respondents \\
\hline
\textit{Age (years)} &       &       \\
Mean    &   54.6    &   54.3    \\
St. Dev.    &   13.4    &   11.5    \\
\hline
\textit{Gender (\%)} &       &       \\
Female  &   66.9    &   68.9    \\
Male    &   33.1    &   31.1    \\
\hline
\textit{Marital status (\%)} &       &       \\
Single  &   10.1    &   $\;$9.8 \\
Married &   79.4    &   80.3    \\
Separated   &   $\;$4.2 &   $\;$3.3 \\
Widowed &   $\;$6.3 &   $\;$6.6 \\
\hline
\textit{Education (\%) } &       &       \\
Primary school  &   12.6    &   12.9    \\
Middle school   &   29.5    &   30.3    \\
High school &   38.8    &   37.6    \\
University  &   19.1    &   19.1    \\
\hline
\textit{Cancer diagnosis (\%) }  &       &       \\
Colon-rectum    &   24.4    &   23.9    \\
Mammary &   45.6    &   46.7    \\
Uterine &   $\;$4.1 &   $\;$3.8 \\
Pulmonary   &   $\;$8.8 &   $\;$8.7 \\
Prostate    &   $\;$4.1 &   $\;$3.8 \\
Other   &   13.0    &   13.0    \\
\hline
Size   &   $\;$275 &   $\;$203 \\
\hline
\end{tabular}}
\end{table}

SF-$36$ is a multidimensional test developed in the
nineties to evaluate HrQoL during the last four weeks of  illness;
it has been validated in many different languages. The test
consists of $36$ polytomous items divided in the
following 9 subsets (corresponding to different latent traits):
\begin{enumerate}
\item PF: physical functioning (10 items);
\item RF: role functioning, that is limitations in daily activities as a result due to physical health problems (4 items);
\item BP: bodily pain (2 items);
\item GH: general health (5 items);
\item VT: vitality (4 items);
\item SF: social functioning (2 items);
\item RE: role-emotional, that is limitations in daily activities as a result due to mental health problems (3 items);
\item MH: mental health (5 items);
\item HC: health change (1 item).
\end{enumerate}

The items have a different number of response categories; to simplify the illustration of the results, in the present
study all the items are dichotomized, with category $1$ indicating the presence of a symptom or limitation
(related to a low level of HrQoL), and category $0$ indicating its absence (related to a high level of the HrQoL).
 Table \ref{tab:dichot} shows, for each item, which original responses correspond to new category 0 and which original responses correspond to new category 1. For the item coding used for the questionnaire SF-$36$ see $[$WAR 02$]$.

\begin{table}[ht]\centering\vspace*{0.5cm}
\caption{\em SF-$36$: original response categories and their dichotomization.} \label{tab:dichot} {\small
\begin{tabular}{lll}
\hline \hline
    & \multicolumn2l{Dichotomized responses} \\
\cline{2-3}
Item*                 &                 \multicolumn1l{0}       &   \multicolumn1l{1}            \\
\hline
1                    &  1, 2, 3                  &     4, 5                        \\
2                    &  1, 2, 3                  &     4, 5                        \\
3a - 3l              &  3                        &     1, 2                      \\
4a - 4d              &  2                        &     1                          \\
5a - 5c              &  2                        &     1                          \\
6                    &  1, 2, 3                  &     4, 5                        \\
7                    &  1, 2, 3                  &     4, 5, 6                     \\
8                    &  1, 2, 3                  &     4, 5                        \\
9a, 9d, 9e, 9h       &  1, 2, 3, 4               &     5, 6                        \\
9b, 9c, 9f, 9g, 9i   &  5, 6                     &     1, 2, 3, 4                  \\
10                   &  3, 4, 5                  &     1, 2                        \\
11a, 11c             &  4, 5                     &     1, 2, 3                    \\
11b, 11d             &  1, 2, 3                  &     4, 5                        \\
\hline
\multicolumn3l{* Item labels are referred to the Italian}\\
\multicolumn3l{version of SF-$36$.}\\
\end{tabular}}\vspace*{0.5cm}
\end{table}

HADS is a  questionnaire designed to assess anxiety and depression
in patients with organic diseases, such as cancer. The
questionnaire is composed by $14$ items referred to two
dimensions:
\begin{enumerate}
\item anxiety (7 items);
\item depression (7 items).
\end{enumerate}
All items have four response categories but have been dichotomized for our analysis, with category $1$
(corresponding to original categories 2 and 3)
indicating the presence of anxiety
(or depression), and category $0$
(corresponding to original categories 0 and 1)
indicating its absence. For the item coding used for the questionnaire HADS see $[$ZIG 83$]$.

Note that, in both questionnaires, every item is assumed to measure
only one latent trait, but these latent traits may be correlated.
\section{The multidimensional latent class Rasch model}\label{sec3}
In the following, we describe the model that we
adopt for the analysis of HrQoL data. We recall that this is a
version of the Rasch model which belongs the class of IRT models
proposed in $[$BAR 07$]$ and that the main
differences with respect to the classic Rasch model $[$RAS 61$]$ are
that (\textit{i})
multidimensionality is assumed
and (\textit{ii}) the latent traits are assumed to have
a discrete distribution. To facilitate the illustration of the model, we make explicit reference to the application described in Section 2.
\subsection{Model assumptions}
Let $n$ denote the number of subjects in the sample and suppose that
they respond to $k$ test items which measure $D$ different latent
traits or dimensions. Also let $\mathcal{I}_d$, $d = 1, \ldots, D$,
be the subset of $\mathcal{I}=\{1, \ldots, k\}$ containing the
indices of the items measuring the latent trait of type $d$ and let
$k_d$ denotes the cardinality of this subset, so
that $k=\sum_{d=1}^Dk_d$. Since we assume that
every item measures only one latent trait, the subsets
$\mathcal{I}_d$ are disjoint; obviously, these
latent traits may be correlated. In our study, $n = 203$ and $k=50$
($36$ items from SF-$36$ and $14$ items from HADS) and, on the basis
of validation studies about these questionnaires, we know that the
items measure $D = 11$ different latent traits: 9 related to HrQoL
and 2 related to psychopathological disturbs.

In this context, the classic Rasch model, based on the assumption
\begin{equation}
\textrm{logit}[p(X_i=1\mid \Theta = \theta)] = \theta - \beta_i,
\quad i = 1, \ldots, k,\label{eq:unid}
\end{equation}
where $\theta$ can be discrete or continuous,  is extended in a
multidimensional way on the basis of the following
parameterization:
\begin{equation}
\textrm{logit}[p(X_i=1\mid \bTheta = \btheta)] = \sum_{d=1}^{D}
(\delta_{id} \theta_d - \beta_i), \quad i = 1, \ldots, k.\label{eq:multid}
\end{equation}
In the above expressions,  $X_i$ is the random variable corresponding
to the response to item $i$; in particular, $X_i=1$ means the
presence  of the symptom or disturb detected by the item and $X_i=0$
means its absence. Moreover, in the unidimensional version,
$\Theta$ is the  latent random variable
corresponding to the trait of the interest and $\theta$ denotes one
of its possible values, whereas $\beta_i$ is  the difficulty of
item $i$. For the multidimensional version, $\bTheta = (\Theta_1,
\ldots, \Theta_{D})\tr$ is the vector of latent variables
corresponding to the different traits measured by the test items,
$\b\theta=(\theta_1,\ldots,\theta_D)\tr$ denotes one
of its possible realizations, and $\delta_{id}$ is a dummy variable
equal to $1$ if item $i$ belongs to $\mathcal{I}_d$ (and then it
measures the $d$th latent trait) and to 0 otherwise.

Note that, as already mentioned, the adopted model
is based on a specific formulation of multidimensionality, in which
each item measures only one latent trait; we are then assuming the
so-called \emph{between-item multidimensionality}. This is
different from the so-called \emph{within-item multidimensionality},
in which each item measures more than one latent trait (see Adams et
al. $[$ADA 97$]$ for a detailed description).
Another important point is that the unidimensional
formulation based on (\ref{eq:unid}) is a special case of
(\ref{eq:multid}). In particular, one of the possible situations in which formulation (\ref{eq:multid}) specialises into  (\ref{eq:unid})  is when the possible ability vectors $\b\th$ have
 elements equal each other.

\begin{figure}[ht!]
    \centering
    \includegraphics[width=\linewidth]{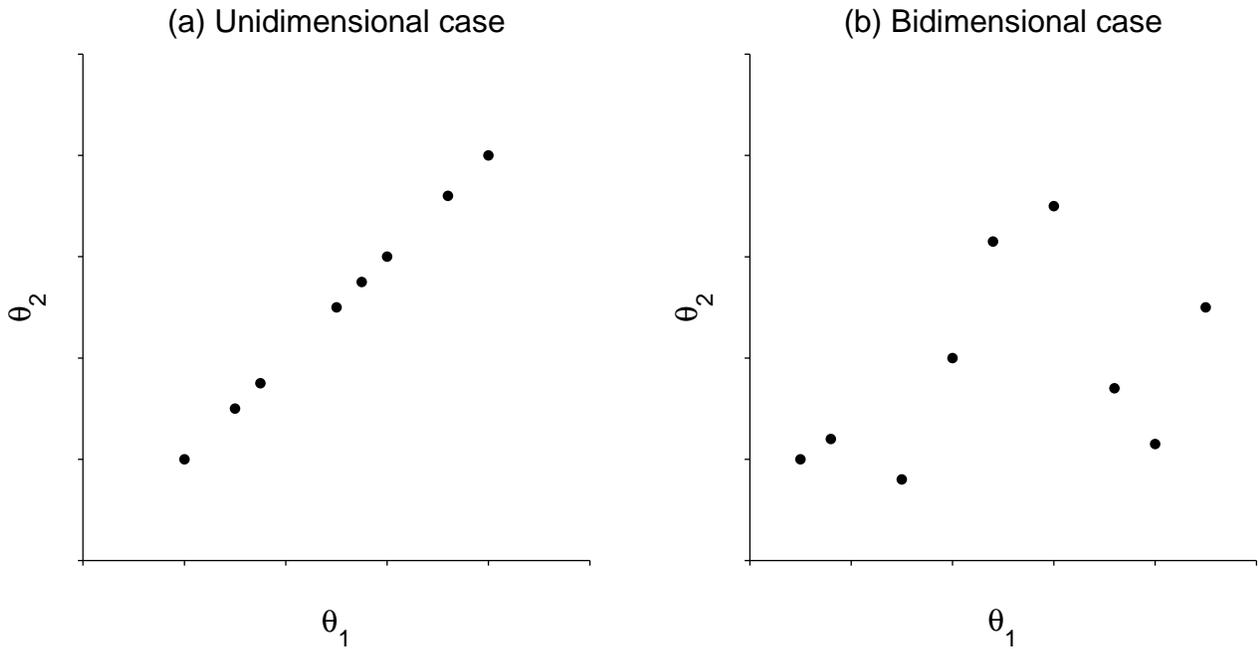}
    \caption{\textit{Plots of ($\theta_1$, $\theta_2$) possible values.}}
    \label{fig1}
\end{figure}

In order to clarify the above point, consider a
simple example involving only two latent traits, so that
$\b\Th=(\Th_1,\Th_2)\tr$, which may assume the values represented in
Figure~\ref{fig1}.

We observe that panel (\textit{a}) corresponds to
the situation of unidimensionality, because all points are aligned
on the bisector. In other words, one latent trait coincides with
the other and the actual number of dimensions is one. On the
contrary, in panel (\textit{b}) the points are not on a bisector
and there is not a complete ordering of them; in this case it is not
possible to completely determine one latent trait from the other and
we can consider the two latent traits as corresponding to two
distinct dimensions.

In order to complete the model specification, we
need to assume a distribution for the vector $\bTheta$, since we
adopt a random-effects rather than a fixed-effects approach. In the
latter, a distinct vector of parameters would be estimated for each
subject, but this would not allow certain
analyses which are of interest in the present context.

In a standard formulation, a continuous
distribution $\pi(\btheta)$, such as the normal distribution, is
adopted. In such a case, the \emph{manifest distribution} of the full
response vector, $\mathbf{X} = (X_1, \ldots, X_{k})\tr$, is given by
\begin{equation}
p(\mathbf{x})=p(\mathbf{X}=\mathbf{x})=\int_{\bl\theta}
p(\mathbf{x}| \btheta) \pi(\btheta)d\btheta, \label{equ:normal}
\end{equation}
where, because of the assumption of \emph{local independence},
\begin{equation}
p(\mathbf{x}| \btheta) = p(\mathbf{X}=\mathbf{x}| \bTheta=\btheta) =
\prod_{i=1}^k p(X_i=x_i \mid \bTheta=\btheta).\label{equ1}
\end{equation}
Note that, because of the between-item multidimensionality
assumption, we assume local independence separately for each
dimension and therefore
\begin{equation}
\prod_{i=1}^k p(X_i=x_i \mid \bTheta=\btheta)= \prod_{d=1}^D
\prod_{h=1}^{k_d} p(X_{dh}=x_{dh} \mid
\bTheta=\btheta)=\prod_{d=1}^D \prod_{h=1}^{k_d} p(X_{dh}=x_{dh}
\mid \Theta_d=\theta_d),
\end{equation}
where $X_{dh}$ denotes the response variable for the $h$-th item
measuring dimension $d$. Note that the notation $X_{dh}$ is equivalent to the previous one, in the sense that each $X_{i}$ coincides with a certain $X_{dh}$ for suitable $d$ and $h$. However, in this way we make explicit the dimension measured by this item.

It is worth recalling that the theory about the
Rasch model does not require the latent trait of interest to have a
continuous distribution. Moreover, assuming a discrete distribution
may have some advantages.
Among others, it can lead to more parsimonious models in given situations, it can
 be used to test the additivity assumption of the Rasch model, and to construct
 empirical Bayes estimators of abilities $[$LIN 91$]$. We adopted this
alternative approach and, therefore, we assume that
the random vector $\bTheta$ has a discrete distribution with support
$\{\bzeta_1, \ldots, \bzeta_C\}$. In other words, we assume that the
population is composed by $C$ latent classes or subpopulations.
We outline that the number of latent classes is the same for each dimension. In theory, it should also be  possible to assume a more general model with a latent class structure  that differs between dimensions, but in this case the interpretability of parameters is not  straightforward.
 As for the conventional LC model (see $[$LAZ 68$]$; $[$GOO 74$]$), the
manifest distribution of the full response vector is  given by
\begin{equation}
p(\mathbf{x})=p(\mathbf{X}=\mathbf{x}) = \sum_{c=1}^C p(\mathbf{x}|
\bzeta_c) \pi_c,\label{equ:1}
\end{equation}
where $\pi_c = p(\bTheta=\bzeta_c)$ is the weight of the $c$th
latent class, that is the probability that a subject belongs to this
latent class. Note that, according to equation (\ref{equ1}),
$p(\mathbf{x}| \bzeta_c)$ is defined as
\begin{equation}
p(\mathbf{x}| \bzeta_c)= p(\b X=\mathbf{x}|\b\Theta=
\bzeta_c)=\prod_{d=1}^D \prod_{h=1}^{k_d} p(X_{dh}=x_{dh} \mid
\Theta_d=\zeta_{cd}), \quad c=1,\ldots,C.
\end{equation}

The specification of the multidimensional LC Rasch model, based on the
assumptions illustrated above, univocally depends on the number of
latent classes ($C$) and on how items are associated to the
different dimensions  (i.e., the subsets $\mathcal{I}_d$, $d =
1,\ldots,D$). Maximum likelihood estimation of this model is
illustrated in the following.
\subsection{Maximum likelihood estimation and model selection}
Assuming that the subjects in the sample are
independent, the log-likelihood of the model is
\begin{equation}
\ell(\boeta)= \sum_{\mathbf{x}}n(\mathbf{x}) \log[p(\mathbf{x})],
\end{equation}
where $\boeta$ is the vector containing all the parameters,  $n(\mathbf{x})$ is the frequency of the response pattern
$\mathbf{x}$, and the probability $p(\mathbf{x})$ is defined as in
equation (\ref{equ:1}) depending on $\boeta$. Moreover, the
sum $\sum_{\mathbf{x}}$ ranges
over all the possible configurations of the
response vector.

In order to maximize $\ell(\boeta)$, and then
obtaining the (discrete) MML estimate of $\b\eta$,  we make use of
the EM algorithm $[$DEM 77$]$, which is an iterative algorithm based
on alternating two steps (E and M). The expectation (E) step
consists of computing the expected value of the
complete log-likelihood evaluated at the current
value of the parameters. The maximization (M) step
consists of updating these
parameters by maximizing the expected
log-likelihood found at the E step. These
parameters are then used to determine the
distribution of the latent variables at the next E step. The two
steps are performed until convergence. See $[$BAR
07$]$ for a detailed description of this estimation algorithm and
its initialization.
At this stage we only stress that, in order to
prevent the problem of the multimodality of the model likelihood, it
is advisable to try different starting values chosen by both
deterministic and random rules. We denote by $\hat{\b\eta}$ the MML
estimate of $\b\eta$, that is the value that, at convergence of EM algorithm, corresponds to the highest log-likelihood. Similar notation is adopted for the single parameters.

Once the model is fitted, we can allocate each subject to one of
the latent classes. The allocation depends on the specific item
response pattern $\mathbf{x}$ provided by the subject. In
particular, for each pattern $\mathbf{x}$ it is possible to estimate
the \emph{posterior probabilities} of belonging to
latent class $c$ as follows
\begin{equation}
\hat{p}(\bzeta_c |\mathbf{x}) = \hat{p}(\bTheta = \bzeta_c
|\mathbf{X}=\mathbf{x}) =
\frac{\hat{p}(\mathbf{x}|\bzeta_c)\hat{\pi}_c}{\sum_{h=1}^C
\hat{p}(\mathbf{x}|\bzeta_h)\hat{\pi}_h}, \quad c=1,\ldots,C.
\end{equation}
An individual is assigned to the latent class with the highest
posterior probability.

An important phase connected with  the estimation process is
represented by the model selection, with special attention to the
choice of the number of latent classes ($C$). A first approach
$[$KIE 56$]$ would suggest to use the value of $C$ corresponding to
the saturation point beyond which the likelihood of the assumed
model fails to increase. For the unidimensional LC Rasch model, the
existence of this saturation point was established by  Lindsay et
al. $[$LIN 91$]$. However, in our context, this number of classes
could be so large that the model becomes non-identifiable or almost
non-identifiable. To avoid this, several authors suggest to use
information criteria, which are typically based on penalizing the
log-likelihood by a factor that takes into account the number of
parameters as a measure of the model complexity.

One of the most known information criteria is
represented by the Bayesian information criterion (BIC) of Schwarz
$[$SCH 78$]$, which is based on the index
\[
BIC = -2\hat{\ell} + g\ln(n),
\]
where $\hat{\ell}$ is the maximum value of the log-likelihood and
$g$ is the  number of non-redundant parameters. For the
multidimensional LC Rasch model, in particular, we have
\begin{equation}
g = (C-1)+DC+(k-D),\label{eq:npar}
\end{equation}
since there are $C-1$ mass probabilities for the classes, $DC$
ability parameters, and $k-D$ difficulty parameters (we have to
consider that some of them are constrained to 0 in order to ensure
model identifiability). Among different models, the one with the
smallest value of the BIC index has to be preferred. For an
illustration in the context of mixture models, which is strongly
related to the present one, see McLachlan and Peel $[$MCL 00$]$, Ch.
6.

\subsection{Software details}
In our application, we used the software
accompanying the paper $[$BAR 07$]$ to estimate the model and choose
the number of classes. This software can be downloaded from the web
page\linebreak \verb"http://www.stat.unipg.it/bartolucci"
and is quite simple to use for practitioners. It is
mainly based on a {\sc Matlab} function, named
\verb"lcest.m", which requires the
 following inputs:
\begin{itemize}
    \item a matrix with each row corresponding to a different
configuration of item responses and each column corresponding to an
item (responses are coded as 0 and 1 and are suitably separated);
    \item a vector with the frequencies of the observed response
configurations;
    \item the number of latent classes;
    \item the type of starting values (deterministic or random);
    \item the type of model (LC model, multidimensional LC Rasch model, multidimesional LC two-parameter logistic model);
    \item a matrix indicating the multidimensional structure of the set of items (any row contains the set of items
    referred to the same dimension).
\end{itemize}

The following output is obtained from
\verb"lcest.m" for the estimated model:
\begin{itemize}
    \item maximum log-likelihood;
    \item estimated probabilities of the latent classes;
    \item BIC index;
    \item estimated difficulty parameters (and discriminating
    indices for the LC two-parameter logistic model);
    \item estimated support points for each latent class and for
    each latent dimension;
    \item estimated probability of responding 1 to every item for each
    latent class;
\item estimated posterior probability of belonging to every latent class for
each response configuration.
\end{itemize}

Function \verb"lcest.m" also
allows us to select the optimal number of latent classes. We have to
estimate a multidimensional LC Rasch model with an increasing number
of latent classes and the best model is chosen on the basis of a
given selection criterion, such as BIC. In a similar way we can
estimate the multidimensional LC Rasch model with different
multidimensional structures in order to choose the optimal number of
dimensions and to evaluate the possibility of collapsing two
dimensions into a single one (see Section 4 for theoretical details
on tests of dimensionality and Section 5 for an application).

Other outputs, such as the correlations between
latent dimensions or between every item and the corresponding
dimension, cannot be obtained directly by \verb"lcest.m",
but a little extra programming is sufficient.
\subsection{Concluding remarks about the model}\label{final}
It is already clear that a crucial assumption of the
multidimensional LC Rasch model described above is that of
discreetness of the latent trait distribution. Alternatively, we
could assume that the vector $\b\Th$ follows a multivariate normal
distribution of dimension $D$. However, if the variance-covariance
matrix of this distribution is unconstrained, computing the manifest
distribution of $\b X$ would require complex numerical tools when
$D$ is greater than 2, as in our application, because a high dimensional integral is involved
(see equation
(\ref{equ:normal})). This has obvious implications on the parameter
estimation and on the possibility of making a comparison with our
model, which is based on a discrete latent trait distribution.

On the other hand, the model based on a multivariate
normal distribution for the latent traits would be easily estimable
under the constraint that the variance-covariance matrix is
diagonal, so that the latent traits are independent. In fact,
estimating this model is equivalent to estimating a separate
unidimensional normal Rasch model for each dimension. The same holds
for the version based on a discrete latent trait distribution. It is
worth noting that, with the same number of latent classes ($C$), the
latter model involves more non-redundant parameters than the
multidimensional LC Rasch model that we propose to use. In fact,
this model has $D(C-1)+DC+(k-D)$ parameters which is larger than the
number of parameters given in (\ref{eq:npar}).

Though estimating separate unidimensional Rasch
models does not allow us to study the correlation between the latent
traits, this may be useful for assessing whether the assumption of
discreteness of the latent trait distribution is suitable or not for
the data at hand. This amount to comparing the global fit of these
models under this assumption with that obtained under the assumption
of normality. Such comparison may be based on standard criteria,
such as BIC.

Finally, we have to remark that the multidimensional
LC Rasch model may be easily extended to deal with polytomous items
and incomplete data. 
Concerning the first issue, the extension
would be based on adopting a suitable parameterization of the
conditional response distribution given the latent traits, which
extends that in (\ref{eq:multid}). At this regard, we can use one of
the parametrizations illustrated by Samejima $[$SAM 96$]$ in her
review of IRT models for polytomous data. About incomplete data, we
have to clarify that the model and the estimation algorithm can be
easily extended to treat the case of incomplete responses by design,
because it is sufficient to drop, for each subject, the items to
which the subject does not respond. The extension to the case of
informative missing responses is not so obvious.
\section{Inference on the correlation between latent traits}\label{sec4}
In our context, the main interest is on the study of the correlation
between the latent traits corresponding to the different dimensions.

First of all, for two dimensions $d_1$ and $d_2$ (e.g., BP and SF),
the correlation may be measured through the index
\begin{equation}
\hat{\rho}_{d_1d_2} = \sum_{c=1}^C
\hat{\zeta}^*_{cd_1}\hat{\zeta}^*_{cd_2} \hat{\pi}_c,\label{param}
\end{equation}
where $\hat{\zeta}_{cd}^*$, $c=1, \ldots, C$, $d=1, \ldots, D$, is
the standardised estimate of the latent trait level  referred to
dimension  $d$ for the subjects in latent class $c$. In practice,
this is the $d$th element of $\hat{\bzeta}_{c}$, once the average
latent trait level has been subtracted from each element and this
difference has been divided by the standard deviation. Moreover,
$\hat{\pi}_c$ is the estimate of the weight of the $c$-th latent
class.

A crucial point is how to test the hypothesis that two dimensions
$d_1$ and $d_2$ are perfectly correlated. This hypothesis is
strongly related to the hypothesis that the items in
$\mathcal{I}_{d_1}$ and $\mathcal{I}_{d_2}$ measure the same latent
trait. Given the parameterization in (\ref{eq:multid}), the latter
may be expressed as $H_0: \zeta_{cd_2}=\zeta_{cd_1} + a$, with $c =
1, \ldots, C$, where $a$ is an arbitrary constant. In the following,
we deal with two different approaches to test $H_0$.

The first approach is directly based on the LR statistic between the
multidimensional latent class Rasch model, in which a separate
latent variable is used to represent each dimension, and a
restricted version of this model in which dimensions $d_1$ and $d_2$
are collapsed into a single one. Both models are estimated by the
discrete MML method, and then by the EM algorithm illustrated at the
end of Section 3. From this estimation we obtain the maximum
log-likelihood of the general model ($\hat{\ell}_{1}$) and that of
the restricted model ($\hat{\ell}_{0}$). It has to be clear that
both models are based on the same number of latent classes; they
only differ in the number of latent variables and in how items are
allocated to these variables. On the basis of these likelihoods, we
obtain the following LR test statistic for the null hypothesis
$H_0$:
\begin{equation}
LR_1 = -2(\hat{\ell}_{0}-\hat{\ell}_{1}).
\end{equation}
Under $H_0$,  $LR_1$ has an asymptotic chi-square distribution with
$C-1$ degrees of freedom, where $C$ is the selected number of
latent classes.  We reject $H_0$ for high values of $LR_1$, that
is values larger than a suitable percentile of the asymptotic
distribution;  otherwise, we do not reject this hypothesis,
implying that the two dimensions are indeed collapsible. In order
to measure the evidence provided by the data in favour of $H_0$,
we can also compute a $p$-value, as the value of the survival
function of the asymptotic distribution at $LR_1$.

One of the main advantages of the above LR test is that, if properly
extended, it can be also used when items discriminate differently
among subjects (i.e., when the Rasch paradigm does not hold);
however, this aspect is beyond the purposes of the present chapter.
On the other hand, this approach requires the choice of the number
of latent classes and the results may depend on this choice. For
this reason we also consider a second approach for testing if two
sets of items measure the same dimension, which does not require to
formulate any assumption on the distribution of the latent
variables. This approach is based on the ML test
$[$GLA 95b$]$. See Martin-L\"{o}f $[$MAR 70$]$ for the original version of the test
for dichotomous items split in two dimensions and
Christensen et al. $[$CHR 02$]$ for a generalization to polytomous
items and to situations with more than two dimensions.
It is worth mentioning that, in the psychometric
literature, alternative tests are available which may be more
powerful in certain situations. We refer, in particular, to the
class of one degree of freedom tests proposed 
by Verhelst $[$VER 01$]$.

ML test is again an LR test between two models.
The main difference with respect to the test based on statistic
$LR_1$ is that the maximum log-likelihoods of the two models under
comparison are obtained from the CML estimates (see $[$RAS 61$]$;
$[$AND 70$]$; $[$AND 72$]$). The maximum log-likelihood of the
general model obtained in this way is denoted by $\tilde{\ell}_{1}$,
whereas that of the restricted model is denoted by
$\tilde{\ell}_{0}$. The resulting LR test statistic is then
\begin{equation}
LR_2 = -2(\tilde{\ell}_{0}-\tilde{\ell}_{1}).
\end{equation}

In particular, the first log-likelihood is obtained as
\begin{equation}
\tilde{\ell}_{1} = \tilde{\ell}_{1c}^{(1)}+\tilde{\ell}_{1c}^{(2)}+\tilde{\ell}_{1m},\label{decomp}
\end{equation}
where $\tilde{\ell}_{1c}^{(1)}$ is the maximum conditional
log-likelihood for the items in $\mathcal{I}_{d_1}$,
$\tilde{\ell}_{1c}^{(2)}$  is the maximum conditional
log-likelihood for the items in $\mathcal{I}_{d_2}$, both obtained
through the CML method, and $\tilde{\ell}_{1m}$ is the maximum
marginal log-likelihood of the multinomial model for the
distribution of the scores. The latter may be expressed as
\begin{equation}
\tilde{\ell}_{1m} = \sum_{r_{1}=0}^{k_{d_1}}
\sum_{r_{2}=0}^{k_{d_2}} n_{r_{1}r_{2}}\log(n_{r_{1}r_{2}}/n),
\end{equation}
where $r_1$ is the test score for the items in the subset $\cg
I_{d_1}$ (e.g., PF), $r_2$ is the test score for the items in the
subset $\cg I_{d_2}$ (e.g., RF), and $n_{r_{1}r_{2}}$ is the
frequency of subjects with scores $r_1$ and $r_2$ at the two
subsets; see also $[$VER 01$]$. We recall that the test score is the
number of items that are responded by category $1$ and that $k_{d_1}$ and $k_{d_2}$ are the number of items from the first
dimension (e.g., PF) and from the second dimension (e.g., RF),
respectively.

A decomposition
similar to that in (\ref{decomp}) holds for $\tilde{\ell}_{0}$. We
have
\begin{equation}
\tilde{\ell}_{0} = \tilde{\ell}_{0c}+\tilde{\ell}_{0m},\label{decomp2}
\end{equation}
where $\tilde{\ell}_{0c}$ is the maximum conditional
log-likelihood for the items in  $\mathcal{I}_{d_1} \cup
\mathcal{I}_{d_2}$, and
\begin{equation}
\tilde{\ell}_{0m} = \sum_{r=0}^{k_{d_1}+k_{d_2}}
n_{r}\log(n_{r}/n).
\end{equation}
In the above expression, $r$ is the test score achieved at the items
in both subsets, the frequency of which is denoted by $n_{r}$. Under
$H_0$, the test-statistic $LR_2$ has an asymptotic chi-square
distribution with $k_{d_1} k_{d_2} -1$ degrees of freedom.

Though the ML test has the advantage of not
requiring the choice of the number of latent classes neither the
formulation of distribution assumptions on the latent variables,
Verhelst $[$VER 01$]$ and Christensen et al. $[$CHR 02$]$ found that
its power is significantly affected by the length of the
questionnaire and it may be disappointingly low
with strongly correlated latent dimensions.
More precisely, it can be verified that the null
distribution may deviate from the asymptotic chi-square distribution
because of the large number of items in comparison with the sample
size. In the psychometric literature there are several proposals to
solve this problem. For example, Christensen and Kreiner $[$CHR
07$]$ proposed a Monte Carlo test procedure for computing
$p$-values. Moreover, another problem connected with the
ML test concerns the fact that this test can only
be used when the Rasch paradigm holds, since otherwise the CML
method cannot be applied.

In order to take into account
 advantages and  disadvantages
of the two tests described in this section, in our  empirical study
we apply both. The LR test based on the discrete MML is calculated
by means of a set of {\sc Matlab} functions which are available upon
request. The ML test is performed by means of {\sc
STATA} as concerning the computation of the maximum conditional
log-likelihoods ($\tilde{\ell}_{0c}$, $\tilde{\ell}_{1c}^{(1)}$,
$\tilde{\ell}_{1c}^{(2)}$), and through a suitable {\sc Matlab}
function as concerning the marginal counterparts
($\tilde{\ell}_{0m}$, $\tilde{\ell}_{1m}$). The same methods could
be applied by alternative statistical softwares, such as {\sc
MULTIRA} of Carstensen and Rost $[$CAR 01$]$, the \%pml {\sc SAS}
macro of Christensen and Bjorner $[$CHR 03$]$ or
{\sc DIGRAM} of Kreiner $[$KRE 03$]$.

Finally, we have to underline that the above
methodology gives valid results provided that the Rasch paradigm
holds for each dimension. In order to check that this is the case,
we have to fit separate Rasch models for each of these dimensions,
and those  that violate this paradigm should not be included in the
analysis.
For this purpose, several testing procedures are available in
the literature (see, among others, $[$LIN 94$]$ and $[$GLA 95a$]$).
These procedures are mainly based on the comparison with the
two-parameter logistic model and with the saturated model by
chi-square or LR statistics, and on infit and outfit statistics.
\section{Application results}
First of all, with reference to the data described
in Section 2, we verified that the Rasch paradigm clearly holds for
only five dimensions; we recall that each dimension corresponds to a
subset of items measuring a specific aspect and these subsets are
defined by the structure of the questionnaires. Three of these
dimensions are referred to HrQoL, that is BP, SF, and VT, whereas
the other two are referred to psychopathological disturbs, that is
anxiety and depression. Therefore, we based our analysis on the
responses to the items referred to these five dimensions, whereas
the items referred to the other ones were discarded. Overall, the data we analysed concern 22 items.

The first step of the analysis was focused on
verifying if a discrete latent structure, on which the LC Rasch
model is based, is indeed more suitable for the data at hand with
respect to a continuous latent structure based on the normal
distribution. As explained in Section \ref{final}, we cannot
directly compare the multidimensional LC Rasch model with its normal
counterpart because of the difficulties in estimating the latter.
However, we can indirectly perform this comparison by estimating
separate undimensional (LC and normal) Rasch models for each
dimension. The results of this comparison are reported in
Table~\ref{tab:0} in terms of maximum log-likelihood and BIC index
for each model.

\begin{table}[ht]\centering\vspace*{0.5cm}
\caption{\em Maximum log-likelihood and BIC index for the
unidimensional Rasch model fitted for each dimension by the MML
method under the assumption that the latent trait has a normal
distribution and under the assumption that is has discrete
distribution with a number of support points (latent classes)
between 1 and 7; for each dimension, in bold are the data referred
to the model with the smallest BIC.} \label{tab:0} {\small
\begin{tabular}{llcrrrrrrr}
\hline \hline
    & &   \multicolumn1c {normal}               & \multicolumn7c {LC Rasch model - number of latent classes ($C$)} \\
\cline{4-10} \multicolumn2c {Dimension}    & \multicolumn1c {Rasch
model}   &    \multicolumn1c {1}  &  \multicolumn1c {2}  &
    \multicolumn1c {3}  &  \multicolumn1c {4}  &  \multicolumn1c {5}  &  \multicolumn1c {6}   &
      \multicolumn1c {7}  \\
\hline\vspace*{-0.35cm}\\
BP  &   $\hat{\ell}$    &   $\;$$\;$-193.4  &   -206.9  &  {\bf -190.5}  &   -190.5  &   -190.5  &   -190.5  &   -190.5  &   -190.5 \\
    &   $BIC$   &   $\;$$\;$$\;$402.7   &   424.4   &   {\bf 402.2}   &   412.8   &   423.5   &   434.1   &   444.8   &   455.4 \\
\hline\vspace*{-0.35cm}\\
SF  &   $\hat{\ell}$    &   $\;$$\;$-157.8  &   -163.1  &   {\bf -147.2}  &   -147.2  &   -147.2  &   -147.2  &   -147.2  &   -147.2 \\
    &   $BIC$   &   $\;$$\;$$\;$331.5   &   336.9   &   {\bf 315.7}   &   326.4   &   337.0   &   347.6   &   358.3   &   368.9 \\
\hline\vspace*{-0.35cm}\\
VT  &   $\hat{\ell}$    &   $\;$$\;$-495.1  &   -511.0  &   -454.6  &   {\bf -446.8}  &   -446.8  &   -446.8  &   -446.8  &   -446.8 \\
    &   $BIC$   &  $\;$1016.7 &   1043.2 &   941.2   &   {\bf 936.2}   &   946.8   &   957.5   &   968.1   &   978.8 \\
\hline\vspace*{-0.35cm}\\
Anxiety &   $\hat{\ell}$    &   $\;$$\;$-671.6  &   -753.5  &   -653.4  &   {\bf -642.7}  &   -641.0  &   -641.0  &   -641.0  &   -641.0 \\
    &   $BIC$   &   $\;$1385.8 &   1544.3 &   1354.8 &   {\bf 1343.9} &   1351.1 &   1361.8 &   1372.4 &   1383.0 \\
\hline\vspace*{-0.35cm}\\
Depression  &   $\hat{\ell}$    &   $\;$$\;$-636.5  &   -701.6  &   -607.7  &   {\bf -602.0}  &   -602.0  &   -602.0  &   -602.0  &   -602.0 \\
    &   $BIC$   &   $\;$1315.5 &   1440.4 &   1263.2 &   {\bf 1262.4} &   1273.0 &   1283.6 &   1294.2 &   1304.9 \\
\hline\vspace*{-0.35cm}\\
Overall & $\hat{\ell}$  &   -2154.4    &   -2336.1    &   {\bf -2053.5}    &   -2029.2    &   -2027.5    &   -2027.5    &   -2027.5    &   -2027.5     \\
 & $BIC$    &   $\;$4452.2 &   4789.2 &   {\bf 4277.1} &   4281.7 &   4331.4 &   4384.7 &   4437.7 &   4491.0 \\
\hline
\end{tabular}}\vspace*{0.25cm}
\end{table}

We observe that the assumption that the distribution
of the latent traits is discrete has to be preferred to the
assumption that this distribution is normal. In fact, for each
dimension, the unidimensional LC Rasch model always attains a higher
maximum log-likelihood than its normal counterpart, while the
minimum BIC index is always smaller. The same happens at
global level (see the last two rows of Table~\ref{tab:0}).

Then, we fitted the multidimensional LC Rasch model
with an increasing number of latent classes. To this regard, in
Table~\ref{tab:1} we report the values of the maximum log-likelihood
and of the BIC index obtained with a number of
latent classes ($C$) from $1$ to $7$.

\begin{table}[ht]\centering\vspace*{0.5cm}
\caption{\em Maximum log-likelihood and BIC index for the
multidimensional latent class Rasch model with a number of latent
classes between 1 and 7; in bold are the data referred to the model
with the smallest BIC.}  \label{tab:1} {\small
\begin{tabular}{lrrrrrrr}
\hline \hline
       & \multicolumn7c {Number of latent classes ($C$)}
       \\\cline{2-8}
    &    \multicolumn1c{1}  & \multicolumn1c{2} &   \multicolumn1c{3} & \multicolumn1c{4}
    & \multicolumn1c{5} & \multicolumn1c{6}  & \multicolumn1c{7} \\\hline
\vspace*{-0.35cm}    \\
$\hat{\ell}$ & -2321.8  &   -2066.7 &   -2016.3 &   -1989.8 &   -1966.9 &   {\bf -1949.3} &   -1937.7 \\
$BIC$         &   4760.6    &   4282.1  &   4213.2  &   4192.1  &   4178.3  &  {\bf 4175.0}  &   4183.7 \\
\hline
\end{tabular}}
\end{table}

According to the adopted selection criterion, we chose $C = 6$
latent classes corresponding to $BIC = 4,175.0$. We
observe that this model has a better fit than the model which is
equivalent to separate unidimensional LC Rasch models. For the
latter we had $4,277.1$ as minimum value of the BIC index; this
value is reached with 2 latent classes (see Table~\ref{tab:0}). For
the multidimensional LC Rasch model with $C = 6$ latent classes, we
obtained the parameter estimates (support points and class weights)
displayed in Table~\ref{tab:4}. In particular, we
estimated 5 support points (corresponding to the number of
dimensions) for each of the 6 latent classes. Due to the adopted
parameterization, see equation (\ref{eq:multid}), very low values of
these support points correspond to a negligible probability of
response $1$, whereas very high values correspond to a negligible
probability of response $0$. To facilitate the interpretation of the
results, the latent classes are ordered on the basis of the
conditional probability of response 1 for the first dimension (BP).

\begin{table}[ht]\centering\vspace*{0.5cm}
\caption{\em Estimated support points and probabilities of the
latent classes (for each dimension, in bold the highest support
point, in italic the smallest one).} \label{tab:4} {\small
\begin{tabular}{lrrrrrr}
\hline \hline
    & \multicolumn6c{Latent class} \\
\cline{2-7}
Dimension   &   \multicolumn1c{1}       &   \multicolumn1c{2}       &   \multicolumn1c{3}   &
\multicolumn1c{4}       &   \multicolumn1c{5}    &   \multicolumn1c{6}           \\
\hline
BP  &   $-\infty$ &   -2.17   &   -0.92   &   -0.91   &   -0.20   &   \textbf{1.48} \\
SF  &   -2.56   &   $-\infty$ &   -1.71   &   -2.92   &  $-\infty$ &   \textbf{0.79} \\
VT  &   -2.36   &   \textit{-3.28}  &   -0.46   &   \textbf{1.09}   &   -2.23   &   0.93 \\
Anxiety &   \textbf{0.00}   &   \textit{-4.87}  &   -3.40   &   -1.85   &   -1.55   &   -0.05 \\
Depression  &   \textbf{1.30}   &   \textit{-4.10}  &   -1.99   &   0.12    &   -1.50   &   0.76 \\
\hline
Probability    & 0.042   &   0.223   &   0.383   &   0.134   &   0.131   &   0.086   \\
\hline
\end{tabular}}\vspace*{0.5cm}
\end{table}

A crucial point of the present study is the interpretation of the
latent classes, which define homogeneous groups in the population.
This interpretation is not always easy, but some
useful suggestions come from the estimates of the support points and
class probabilities. In particular, the smallest classes are the
first (4.2\% of subjects) and the last (8.6\% of subjects). These
classes contain, respectively, patients with the highest tendency to
psychopathological disturbs and with the highest impairment with
respect to BP and SF. The second class contains more subjects
($22.3\%$); these subjects are in the best conditions with respect
all dimensions, with the exception of BP. The third class is the
largest ($38.3\%$ subjects) and includes subjects with an
intermediate tendency to be ill with respect to each dimension.
Finally, the fourth and fifth latent classes have a very similar
size (13.4\% and 13.1\% of subjects) and include, respectively,
subjects with the worst conditions with respect to VT and the best
conditions with respect to SF.
Note that individuals belonging to the
second and the fifth latent class show a very good level with respect to SF.

As already mentioned, it is  important to analyze the
correlation between the different dimensions of HrQoL, and the
correlation with anxiety and depression. At this aim, in
Table~\ref{tab:3} we report the correlation coefficients estimated
on the basis of equation (\ref{param}) for every pair of dimensions.

\begin{table}[ht]\centering\vspace*{0.5cm}
\caption{\em Estimated correlation matrix between latent traits.}
\label{tab:3} {\small
\begin{tabular}{lcccc}
\hline \hline
     &   SF  &   VT  &   Anxiety &   Depression  \\
\hline
BP     &   0.679   &   0.794   &   0.932   &   0.989   \\
SF  &          &   0.934   &   0.388   &   0.596   \\
VT  &       &          &   0.530   &   0.707   \\
Anxiety &       &       &          &   0.971   \\
\hline
\end{tabular}}\vspace*{0.5cm}
\end{table}

As regards the dimensions of  HrQoL, we observe that VT is strongly
correlated  especially with the other psychological dimension SF
($\hat{\rho} = 0.934$). As regards the association between the
dimensions of HrQoL measured by the SF-$36$ questionnaire and the
psychopathological disorders measured by the HADS questionnaire,
both anxiety and depression show  a very high correlation with BP
($\hat{\rho}=0.932$ and $\hat{\rho}=0.989$, respectively). On the
other hand, the association of anxiety with SF and VT is much weaker
($\hat{\rho}=0.388$ and $\hat{\rho}=0.530$, respectively), and the
association of depression with SF and VT assumes intermediate values
($\hat{\rho}=0.596$ and $\hat{\rho}=0.707$, respectively). Finally,
as we may expect, the correlation between anxiety and depression is very
high ($\hat{\rho}=0.971$).

When the correlation between two latent traits is high, it is
interesting to verify if the corresponding items actually measure
the same dimension. At this aim, we performed both tests described
in the previous section, that is the LR tests based on the discrete
MML and the CML methods, for every pair of dimensions. LR statistics
and the corresponding $p$-values are shown in Table~\ref{tab:3a} for
the first test and in Table~\ref{tab:3b} for the second one.

\begin{table}[ht]\centering\vspace*{0.5cm}
\caption{\em Results from the test of dimensionality based on the
discrete MML parameter estimates; in brackets the $p$-values.}
\label{tab:3a} {\small
\begin{tabular}{lrrrr}
\hline \hline
    &   \multicolumn1c {SF} &   \multicolumn1c {VT} &   \multicolumn1c {Anxiety}    &   \multicolumn1c {Depression} \\
\hline
BP  &   27.1 (0.000)    &   14.0 (0.007)    &   22.2 (0.000)    &   27.5 (0.000)    \\
SF  &       &   12.5 (0.014)    &   32.1 (0.000)    &   26.3 (0.000)    \\
VT  &       &       &   51.2 (0.000)    &   70.3 (0.000)    \\
Anxiety &       &       &       & 9.1 (0.059)   \\
\hline
\end{tabular}}
\end{table}

\begin{table}[ht]\centering
\caption{\em Results from the test of dimensionality based on the
CML parameter estimates (ML test); in brackets the $p$-values.} \label{tab:3b}
{\small
\begin{tabular}{lrrrr}
\hline \hline
    &   \multicolumn1c {SF} &   \multicolumn1c {VT} &   \multicolumn1c {Anxiety}    &   \multicolumn1c {Depression} \\
\hline
BP  &   23.0 (0.000)    &   38.6 (0.000)    &   53.7 (0.000)    &   57.0 (0.000)    \\
SF  &       &   30.7 (0.000)    &   42.9 (0.000)    &   36.3 (0.001)    \\
VT  &       &       &   126.7 (0.000)   &   112.7 (0.000)   \\
Anxiety &       &       &       &   83.3 (0.001)    \\
\hline
\end{tabular}}\vspace*{0.5cm}
\end{table}

We observe that the values of the test statistic based on discrete
MML method are almost always smaller than the corresponding values
computed on the basis of the CML method, but the main conclusions
are similar. In fact, both test statistics have a
null asymptotic chi-square distribution, but the second is based on
a larger number of degree of freedom. Nevertheless, both kinds of
test do not show any evidence of unidimensionality. A weak evidence
of unidimensionality appears under the test based on the discrete
MML estimates for anxiety and depression ($p$-value equal to
$0.059$) and for VT and SF ($p$-value equal to $0.014$). Then, the
present study confirms that the dimensions defined by the SF-$36$
and HADS questionnaires are indeed separate, and collapsing some of
them would imply an inappropriate simplification of the phenomena
investigated by the test items.

\section*{Acknowledgements}

The authors are grateful to A. Bonacchi of University of Florence (IT) for making available the data. F. Bartolucci acknowledges the financial support of PRIN07 (grant 2007XECZ7L003) and of the ``Einaudi Institute for Economics and Finance'', Rome, IT.

\section*{References}
{\small\begin{description}
\item $[$AND 70$]$ ANDERSEN E.B., Asymptotic properties of conditional maximum likelihood estimators, {\em Journal of the Royal Statistical Society, Series B}, vol. 32, 1970, p. 283-301.\vspace*{-0.2cm}
\item $[$AND 72$]$ ANDERSEN E.B., The numerical solution of a set of conditional estimation equations, {\em Journal of the Royal Statistical Society, Series B}, vol. 34, 1972, p. 42-54.\vspace*{-0.2cm}
%
%
\item $[$ADA 97$]$ ADAMS R.J., WILSON M., WANG W.C., The multidimensional random coefficients multinomial logit model, {\em Applied Psychological Measurement}, vol. 21, 1997, 1-23.\vspace*{-0.2cm}
\item $[$BAR 07$]$ BARTOLUCCI F., A class of multidimensional {IRT} models for testing unidimensionality and clustering items, {\em Psychometrika}, vol. 72, 2007, p. 141-157.\vspace*{-0.2cm}
\item $[$CAR 01$]$ CARSTENSEN C.H., ROST J., {MULTIRA} (version
1.65) [Computer software and manual], 2001, retrieved from
http://www.multira.de.\vspace*{-0.2cm}.
\item $[$CHR 02$]$ CHRISTENSEN K.B., BJORNER J.B., KREINER S.,
PETERSEN J.H., Testing unidimensionality in polytomous Rasch models.
{\em Psychometrika}, vol. 67 no.4, 2002, p. 563-574.
\vspace*{-0.2cm}
\item $[$CHR 03$]$ CHRISTENSEN K.B., BJORNER J.B., {SAS} macros
for {R}asch based latent variable modelling (Res. Rep. No. 03/13.
Department of Biostatistics. University of Copenhagen, 2003. \\
Retrieved from
\verb"http://www.ifsv.ku.dk/afdelinger/biostatistik/lokale/sasmacros".\vspace*{-0.2cm}
\item $[$CHR 07$]$ CHRISTENSEN K.B., KREINER S., A Monte Carlo
 approach to unidimensionality testing in polytomous Rasch
models. {\em Applied Psychological Measurement}, vol. 31, 2007, p.
20-30. \vspace*{-0.2cm}
\item $[$DEM 77$]$ DEMPSTER A.P., LAIRD N.M., RUBIN D.B., Maximum likelihood from incomplete data via the {EM} algorithm (with discussion). {\em Journal of the Royal Statistical Society, Series B}, vol. 39, 1977, p. 1-38.\vspace*{-0.2cm}
\item $[$FAY 00$]$ FAYERS P.M., MACHIN D., \emph{Quality of life. Assessment, analysis and interpretation}, Wiley, 2000.
\item $[$FOR 95$]$ FORMANN A.K., Linear logistic latent class analysis and the Rasch model, in G.H. Fischer, and I.W.Molenaar (Eds.),
\emph{Rasch models: Foundations, recent developments, and applications}, New York, Springer-Verlag, 1995, p. 239-255.
\item $[$GLA 95a$]$ GLAS C.A.W., VERHELST N.D. Testing the Rasch model, in G.H. Fischer, and I.W. Molenaar
(Eds.), \emph{Rasch models: Foundations, recent developments, and
applications}, New York, Springer-Verlag, 1995, p. 69-95.
\item $[$GLA 95b$]$ GLAS C.A.W., VERHELST N.D. Tests of fit for
polytomous Rasch models, in G.H. Fischer, and I.W. Molenaar
(Eds.), \emph{Rasch models: Foundations, recent developments, and
applications}, New York, Springer-Verlag, 1995, p. 325-352.
\item $[$GOO 74$]$ GOODMAN L.A., Exploratory latent structure analysis using both identifiable and unidentifiable models. {\em Biometrika}, vol. 61, 1974, p. 215-231.\vspace*{-0.2cm}
\item $[$HAR 04$]$ HARDOUIN J.B., MESBAH M., Clustering binary variables in subscales using an extended Rasch model and Akaike information criterion.
{\em Communication in Statistics. Theory and Methods}, vol. 33, 2004, p.
1277-1294.\vspace*{-0.2cm}
\item $[$KIE 56$]$ KIEFER J., WOLFOWITZ J., Consistency of the
maximum likelihood estimator in the presence of infinitely many
incidental parameters. {\em Annals of Mathematical Statistics}, vol. 27, 1956, p. 887-906.\vspace*{-0.2cm}
 \item $[$KRE 03$]$ KREINER S.,
\emph{Introduction to {DIGRAM}}, Res.Rep.No. 03/10, Department of
Biostatistics. University of Copenhagen, 2003.
 \item $[$LAZ 68$]$ LAZARSFELD P.F., HENRY N.W.,
\emph{Latent structure analysis}, Boston, Houghton Mifflin, 1968.
\item $[$LIN 94$]$ LINACRE, J. M., and B. D. WRIGHT. 1994. (Dichotomous mean-square) infit and outfit chi-square
fit statistics. Rasch Measurement Transactions 8(2): 360.
\item $[$LIN 91$]$ LINDSAY B., CLOGG C., GREGO J., Semiparametric
estimation in the Rasch model and related exponential response
models, including a simple latent class model for item analysis.
{\em Journal of the American Statistical Association}, vol. 86, 1991, p.
96-107.\vspace*{-0.2cm}
\item $[$MAG 01$]$ MAGIDSON J., VERMUNT J.K., Latent class factor
and cluster models, bi-plots, and related graphical displays. {\em Sociological Methodology}, vol. 31, 2001, p. 223-264.\vspace*{-0.2cm}
\item $[$MCL 00$]$ MCLACHLAN G. J., PEEL, D., {\em Finite Mixture Models},
Wiley, 2000.
\item $[$MAR 70$]$ MARTIN-L\"{O}F P., {\em
Statistika modeller} (Statistical Models): Anteckningar fr\.{a}n seminarier las\.{a}ret 1969-1970 (Notes from seminars in the academic year 1969-–1970), with the assistance of  {R}olf {S}undberg.  Stockholm: Instit\"{u}tet f\"{o}r F\"{o}rs\"{a}kringsmatemetik och Matematiks Statistiks vid Stockholms Universitet, 1970.\vspace*{-0.2cm}
\item $[$MES 10$]$ MESBAH M., Statistical Quality of Life, in N. Balakrishnan (Ed.),
\emph{Methods and Applications of Statistics in Life and Health Sciences}, New York, Wiley, 2010, p. 839-864.
%
\item $[$RAS 61$]$ RASCH G., On general laws and the meaning of measurement in psychology, \emph{Proceedings of the IV Berkeley Symposium on Mathematical Statistics and Probability}, vol. 4, 1961, p. 321-333.\vspace*{-0.2cm}
\item $[$ROS 90$]$ ROST J., Rasch models in latent classes: an integration of two approaches to item analysis, \emph{Applied Psychological Measurement}, vol. 14, 1990, p. 271-282.\vspace*{-0.2cm}
\item $[$SAM 96$]$ SAMEJIMA F., Evaluation of mathematical models for ordered polychotomous responses.
{\em Behaviormetrika}, vol. 23, 1996, p. 17–-35.\vspace*{-0.2cm}
\item $[$SCH 78$]$ SCHWARZ G., Estimating the dimension of a model.
{\em Annals of Statistics}, vol. 6, 1978, p. 461–-464.\vspace*{-0.2cm}
\item $[$WAR 02$]$ WARE J.E., KOSINSKI M., DEWEY J.E., \emph{How to Score Version 2 of the SF-36 Health Survey (Standard \& acute forms)}. Lincoln RI, Quality Metric Incorporated, 2002.\vspace*{-0.2cm}
\item $[$VER 01$]$ VERHELST, N.D., Testing the unidimensionality assumption of the Rasch model,
{\em Methods of Psychological Research Online}, vol. 6, 2001, p. 231–-271.\vspace*{-0.2cm}
\item $[$WHO 95$]$ The WHOQOL Group, The {W}orld {H}ealth {O}rganization {Q}uality of {L}ife Assessment ({WHOQOL}): position paper from the {W}orld {H}ealth {O}rganization, {\em Special Issue on Health-Related Quality of Life: what is it and how should we measure it? Social Science and Medicine}, vol. 41(10), 1995, p. 1403-1409.\vspace*{-0.2cm}
\item $[$ZIG 83$]$ ZIGMOND A.S., SNAITH R.P., The hospital anxiety and depression scale, \emph{Acta Psychiatrika Scandinavica}, vol. 67 no. 6, 1983 p. 361-370.\vspace*{-0.2cm}
\end{description}}
\end{document}